\documentclass[aps,pra,superscriptaddress,twocolumn]{revtex4-1}
\usepackage{bbm}
\usepackage{mathrsfs}
\usepackage{amsmath}
\usepackage{amsfonts}
\usepackage[colorlinks=true,linkcolor=blue,urlcolor=blue,citecolor=blue,anchorcolor=blue]{hyperref}
\usepackage{graphicx,epstopdf}
\usepackage{subfigure}
\usepackage{epsfig}
\usepackage{dcolumn}
\usepackage{bm}
\usepackage{color}
\usepackage{natbib}
\usepackage{amssymb}
\usepackage{xcolor}
\usepackage{braket}
\usepackage{ulem}
\usepackage{float}
\usepackage{lipsum}

\begin{document}
\title{Floquet topological phases with large winding number}

\author{Kaiye Shi}
\affiliation{Department of Physics and Beijing Key Laboratory of Opto-electronic Functional Materials and Micro-nano Devices, Renmin University of China, Beijing 100872, China}
\affiliation{Key Laboratory of Quantum State Construction and Manipulation (Ministry of Education), Renmin University of China, Beijing 100872,China}

\author{Xiang Zhang}
\email{siang.zhang@ruc.edu.cn}
\affiliation{Department of Physics and Beijing Key Laboratory of Opto-electronic Functional Materials and Micro-nano Devices, Renmin University of China, Beijing 100872, China}
\affiliation{Key Laboratory of Quantum State Construction and Manipulation (Ministry of Education), Renmin University of China, Beijing 100872,China}
\address{Beijing Academy of Quantum Information Sciences, Beijing 100193, China}

\author{Wei Zhang}
\email{wzhangl@ruc.edu.cn}
\affiliation{Department of Physics and Beijing Key Laboratory of Opto-electronic Functional Materials and Micro-nano Devices, Renmin University of China, Beijing 100872, China}
\affiliation{Key Laboratory of Quantum State Construction and Manipulation (Ministry of Education), Renmin University of China, Beijing 100872,China}
\address{Beijing Academy of Quantum Information Sciences, Beijing 100193, China}

\begin{abstract}

Recently, anomalous Floquet topological phases without static counterparts have been observed in different systems, where periodically driven models are realized to support a winding number of 1 and a pair of edge modes in each quasienergy gap. Here, we focus on cold atomic gases in optical lattices and propose a novel driving scheme that breaks rotation symmetry but maintains inversion symmetry of the instantaneous Hamiltonian, and discover a novel type of anomalous Floquet topological phase with winding number larger than 1. By analyzing the condition of band touching under symmetry constraint, we map out the phase diagram exactly by varying the driving parameters and discuss the quasienergy spectra of typical topological phases, which can present multiple pairs of edge modes within a single gap. Finally, we suggest to characterize the topology of such phases by detecting the band inversion surfaces via quench dynamics. 
\end{abstract}

\maketitle

\section{Introduction}
\label{sec:intro}

Since the discovery of quantum Hall effect~\cite{Klitzing1980} with quantized Hall resistance plateaus in two-dimensional (2D) electron systems in 1980, the study of topological properties of condensed matter systems have aroused extensive interests. This wave of research leads to the discovery of versatile topological systems, including topological insulators~\cite{Fu2007,Goldman2010,Hasan2010,Qi2011}, topological superconductors~\cite{Qi2011,Leijnse2012,Sato2017}, Weyl semimetals~\cite{Burkov2011,Xu2015} and nodal line semimetals~\cite{Mullen2015,Weng2015}, etc. The topological nature of such states is characterized by topological invariants. For a non-interacting 2D lattice system without time reversal symmetry, the topological invariant is the well-known Chern number~\cite{Thouless1982}, which not only captures the topology of bulk spectrum, but also corresponds to the net number of one-dimensional topologically protected edge modes at system boundaries, i.e., the bulk-boundary correspondence~\cite{Hatsugai1993}. Recently, the exploration of topological phases with large Chern number (higher than 1) has attracted much attention~\cite{Wang2013,Skirlo2015}. On one hand, it is suggested that a fractional filling of topological phase with large Chern number will lead to new topological states~\cite{Qi2012,Wang2012}. On the other hand, a realization of such large-Chern-number phase can improve the performance of topological quantum devices and may be used for multi-channel quantum computing. Up to date, several topological materials with large Chern number have been reported in static systems at thermal equilibrium~\cite{Zhao2020,Felser2020}.

\begin{figure}[tbp]
	\centering
	\includegraphics[width=1.0\linewidth]{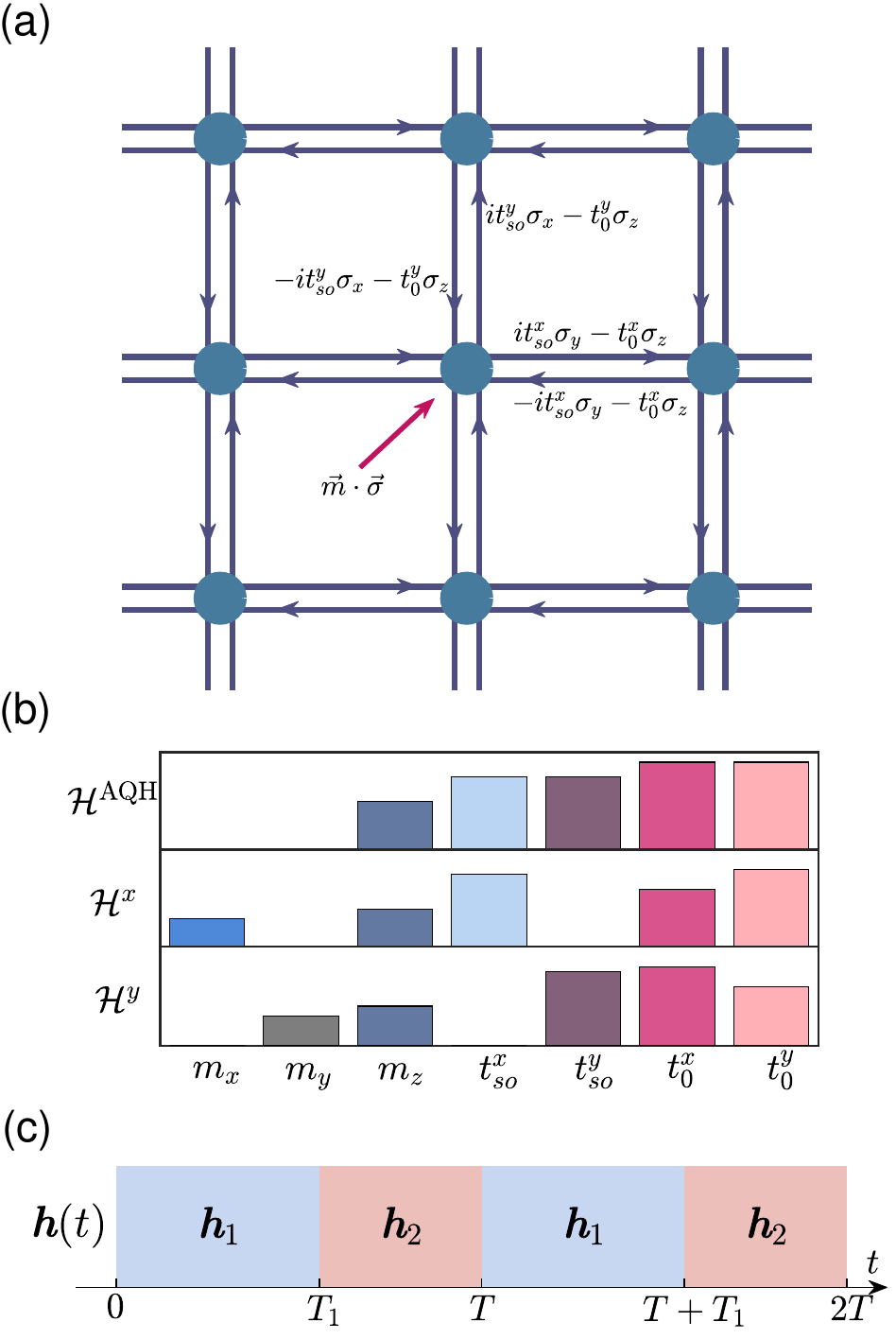}
	\caption{(a) Illustration of the 2D tight-binding model with nearest-neighbor hopping and spin-orbit coupling (blue arrows), and on-site magnetic field (red arrow). (b) Parameter selection for different types of Hamiltonians, ${\cal H}^{\mathrm{AQH}}$, ${\cal H}^x$ and ${\cal H}^y$. (c) The two-step quench protocol with period $T$.}
	\label{fig1}
\end{figure}

Another intriguing direction for the search of exotic phases is Floquet engineering, where a system is subjected to periodical driving and stroboscopic measurement. Novel topological phenomena without static counterpart have been predicted~\cite{Goldman2014,Eckardt2017,Rudner2013,Rudner2020,Liu2019,sky2022}, such as the anomalous Floquet topological (AFT) phase~\cite{Rudner2013,Rudner2020}, where the Chern numbers of all quasienergy bands are zero but robust edge modes exist in gaps. A modified bulk-boundary correspondence is established with winding number defined in the time-momentum space~\cite{Rudner2013} to characterize the topology of the quasienergy gaps. Experimental simulation of such exotic anomalous Floquet topological systems can be found in acoustic systems~\cite{Peng2016}, photonic crystals~\cite{Maczewsky2017,Mukherjee2017}, and cold atomic gases~\cite{Bloch2020}.

Among these physical platforms, the quantum nature and high controllability of atomic gas make it a powerful candidate for studying Floquet topological phases. Many theoretical~\cite{Eckardt2017,Liu2019,sky2022} and experimental progresses~\cite{Bloch2020,Zhang2023} have been reported in this direction. Notable examples include the AFT phase with large Chern number~\cite{Liu2019}, where winding numbers of different gaps are opposite to each other, and the type-II AFT phases~\cite{sky2022, Zhang2023}, which cannot be characterized by winding numbers. In all these cases, a general principle if established, stating that the net number of edge states in a quasienergy gap corresponds to the winding number associate with that gap~\cite{Rudner2013}. 
In that sense, the winding number in a Floquet system plays the equivalent role as the Chern number in a static system. Therefore, the search for topological quantum phases with large winding number in periodically driven systems has great value in both conceptual and application aspects, but has not yet been discovered.

In this manuscript, we propose new driving schemes for a 2D two-band model realized by cold atomic gas with synthetic spin-orbit coupling in a square Raman lattice~\cite{Wu2016,Sun2018,S-Y2018,Yi2019,Liang2023}, and find topological phases with winding number $\pm 2$. Compared with the scheme to achieve Floquet topological phase with large Chern number~\cite{Liu2019}, our driving protocol breaks the $C_4 $ rotational symmetry~\cite{Wu2016,Sun2018} and consequently induces more band touching points. By tuning the driving parameters and mapping out the phase diagram, we find multiple Floquet topological phases with large winding numbers and multiple pairs of robust edge modes in a single gap. Finally, we propose to use the stroboscopic time-averaged spin textures in quench dynamics to characterize the topological nature of different phases.

\section{Two-dimensional Raman lattice under periodic driving}
\label{sec:model}

We consider a non-interacting 2D model on a square lattice with spin-orbit coupling and on-site magnetic field, as illustrated in Fig.~\ref{fig1}(a). This model has been realized by spin-orbit coupled ultracold atoms trapped in a square optical Raman lattice~\cite{Wu2016,Sun2018,S-Y2018,Yi2019,Liang2023}. Within tight-binding approximation, the Bloch Hamiltonian $\mathcal{H}(\bm{q})=\bm{h}(\bm{q})\cdot \bm{\sigma}$ can be expressed as~\cite{Yi2019}
\begin{equation}
	\label{eq1}
	\begin{aligned}
		\mathcal{H}(\bm{q})=&(m_x+2t^y_{so}\sin{q_y})\sigma_x+(m_y+2t^x_{so}\sin{q_x})\sigma_y\\
		&+(m_z-2t^x_0\cos{q_x}-2t^y_0\cos{q_y})\sigma_z,
	\end{aligned}
\end{equation}
where $\bm{q}=(q_x,q_y)$ is the quasimomentum, $t^j_0$ ($t^j_{so}$) denotes the spin-conserved (spin-flip) hopping coefficient with $j=x,y$ the spatial dimensions, $m_i$ is the Zeeman constant, and $\sigma_i$ is the Pauli operator with $i=x,y,z$ in spin space. A special example of this Hamiltonian has been discussed in Ref.~\cite{Wu2016,Sun2018,S-Y2018} with $m_x=m_y=0$, $t^x_0=t^y_0=t_0$ and $t^x_{so}=t^y_{so}=t_{so}$, such that the model reduces to an anomalous quantum Hall (AQH) model
\begin{eqnarray}
\mathcal{H}^{\mathrm{AQH}}(\bm{q})&=&\bm{h}^{\mathrm{AQH}}(\bm{q})\cdot \bm{\sigma}
\nonumber \\
&=& 2t_{so}\sin{q_y}\sigma_x+2t_{so}\sin{q_x}\sigma_y
\nonumber \\
&&+(m_z-2t_0\cos{q_x}-2t_0\cos{q_y})\sigma_z.
\end{eqnarray}
The topological nature of this static system is characterized by the first Chern number $C_n=\frac{i}{2\pi}\int_{\mathrm{FBZ}} dq_xdq_y\mathrm{Tr}(P_n[\partial_{q_x}P_n,\partial_{q_y}P_n])$, where $P_n(\bm{q})=\ket{\psi_n(\bm{q})}\bra{\psi_n(\bm{q})}$ and $\ket{\psi_n(\bm{q})}$ are eigenstates of the $n$-th band, and the integration is conducted within the first Brillouin zone (FBZ)~\cite{Sticlet,Avron}. By changing the relative phase of the two orthogonally polarized components of the laser~\cite{Yi2019}, two other topologically trivial systems can be obtained with effective Hamiltonians 
\begin{eqnarray}
\mathcal{H}^{x}(\bm{q}) &=& \bm{h}^{x}(\bm{q})\cdot \bm{\sigma}
\nonumber \\
&=&
m_x\sigma_x+2t^x_{so}\sin{q_x}\sigma_y 
\nonumber \\
&&
+
(m_z-2t^x_0\cos{q_x}-2t^y_0\cos{q_y})\sigma_z, \nonumber \\
\mathcal{H}^{y}(\bm{q}) &=& \bm{h}^{y}(\bm{q})\cdot \bm{\sigma}
\nonumber \\
&=&
m_y\sigma_y + 2t^y_{so}\sin{q_y}\sigma_x
\nonumber \\
&&
+(m_z-2t^x_0\cos{q_x}-2t^y_0\cos{q_y})\sigma_z.
\end{eqnarray}
Typical choices of parameters for the three cases are schematically shown in Fig.~\ref{fig1}(b).

For a time-dependent Hamiltonian with period $T$, the long-time dynamics is described by a periodic evolution operator (with $\hbar=1$) $U_T=\mathbb{T}\ \mathrm{exp}[-i\int^{T}_0H(t)dt]$ with $\mathbb{T}$ the time-ordered operator. The quasienergy $\varepsilon_F$ of the periodically driven system is given by the  the eigenvalue of the effective Floquet Hamiltonian ${\cal H}_F$, which is defined by $U_T$ as ${\cal H}_F \equiv i\log{U_T}/T$. The nontrivial winding of the quasienergy spectrum will lead to the appearance of edge states. However, since the quasienergy spetrum is repeated in energy with period $2\pi/T$, the Chern number alone is not enough to determine the number and chirality of the edge states. For instead, a new topological invariant, the winding number $W$, is introduced as~\cite{Rudner2013}
\begin{equation}
	\label{eq3}
	W_{s}=\int\frac{dtdq_x dq_y}{8\pi^2}\mathrm{Tr} \left(u_{s}^{-1}\partial_{t}u_{s} \left[u_{s}^{-1}\partial_{q_x}u_{s},u_{s}^{-1}\partial_{q_y}u_{s} \right] \right),
\end{equation}
where $s=0$ or $\pi$ labels the gap in quasienergy spectrum, and the modified time-evolution operator $u_{s}(\bm{q},t)$ is 
\begin{equation}
	\label{eq:u}
	u_{s}(\bm{q},t)=\left\{
	\begin{aligned}
		&U(\bm{q},2t),  &t\in[0,T/2),\\
		&e^{-i\mathcal{H}_{s}(\bm{q})2(T-t)},  &t\in [T/2,T),\\
	\end{aligned}
	\right.
\end{equation}	 
with the bulk time-evolution operator $U(\bm{q},t)$ and $\mathcal{H}_{s}(\bm{q}) = \frac{i}{T}\log_{-s}[U(\bm{q},T)]$. Notice that the $s$ dependence of $u_s$ in Eq.~(\ref{eq3}) roots from the choice of branch cut $e^{-is}$ of the logarithmic function $\log_{-s}$ in the definition of $\mathcal{H}_{s}$~\cite{Rudner2013}.

If one drives the parameters of a topologically trivial system periodically, it is possible to obtain an effective Floquet Hamiltonian with non-trivial topological nature. For example, in the AFT phase~\cite{Peng2016,Maczewsky2017,Mukherjee2017,Bloch2020}, the winding numbers $W_0$ and $W_{\pi}$ associated respectively with the gaps at $\varepsilon=0$ (0 gap) and $\varepsilon=\pi/T$ ($\pi$ gap) are both 1 (or $-1$), leading to quasienery bands with zero Chern number but topologically protected edge states. Alternatively, if one drives the parameters of a topological system, a new type of topological phase can be generated, featuring opposite winding numbers $W_0 = - W_\pi = \pm 1$ in 0 and $\pi$ gaps~\cite{Liu2019,sky2022,Zhang2023}. This phase, later referred as large-Chern-number AFT phase, thus acquires a Chern number of $\pm 2$, and the edge modes in the 0 and $\pi$ gaps possess opposite chirality. However, since this driving scheme and the resulting Floquet Hamiltonian inherit the $C_4$ symmetry of the original static system, the winding of the quasienergy spectrum is restricted such that the absolute value of winding numbers cannot exceed 1.

For simplicity, we consider a two-step quench protocol [Fig.~\ref{fig1}(c)]
\begin{equation}
	\label{eq4}
	\bm{h}(\bm{q},t)=\left\{
	\begin{aligned}
		\bm{h}_1(\bm{q})=&(m_{x1}+2t^y_{so1}\sin{q_y},m_{y1}+2t^x_{so1}\sin{q_x},\\
		&m_{z1}-2t^x_{01}\cos{q_x}-2t^y_{01}\cos{q_y}),\\
		&\ \ t\in [mT,mT+T_1);\\
		\bm{h}_2(\bm{q})=&(m_{x2}+2t^y_{so2}\sin{q_y},m_{y2}+2t^x_{so2}\sin{q_x},\\
		&m_{z2}-2t^x_{02}\cos{q_x}-2t^y_{02}\cos{q_y}),\\
		&\ \ t\in[mT+T_1,(m+1)T),
	\end{aligned}
	\right.
\end{equation}
where $T_1 < T$ is within one period, and $m$ is an integer. This simple form of driving is chosen to give analytic results of the effective Floquet Hamiltonian and the band-touching conditions. For a general choice of driving protocol, one can rely on numerical simulation to obtain qualitatively similar conclusions. 

Depending on the value of parameters, the instantaneous Hamiltonians $\bm{h}_1$ and $\bm{h}_2$ can be chosen from the three types of $\mathcal{H}^{\mathrm{AQH}}$, $\mathcal{H}^x$ and $\mathcal{H}^y$. 
Using the notation of $\bm{h}_1$-$\bm{h}_2$, the driving protocol of Eq.~(\ref{eq4}) can be categorized as: 
\begin{eqnarray}
\begin{array}{cl}
(\mathbb{I}) & \mathcal{H}^x\textrm{-}\mathcal{H}^x, \textrm{or\ } \mathcal{H}^y\textrm{-} \mathcal{H}^y, \\ 
(\mathbb{II}) & \mathcal{H}^{\mathrm{AQH}}\textrm{-}\mathcal{H}^{\mathrm{AQH}},\\
(\mathbb{III}) &\mathcal{H}^x\textrm{-}\mathcal{H}^y,\\
(\mathbb{IV}) & \mathcal{H}^y\textrm{-}\mathcal{H}^{\mathrm{AQH}}, \textrm{or\ }  \mathcal{H}^x\textrm{-}\mathcal{H}^{\mathrm{AQH}}. 
\end{array}
\nonumber
\end{eqnarray}
The Floquet Hamiltonian obtained by class-$\mathbb{I}$ driving is topologically trivial. The conventional AFT phase (with Chern number $0$) and the large-Chern-number AFT phase (with Chern number $\pm 2$) can be realized in class-$\mathbb{III}$~\cite{Bloch2020} and class-$\mathbb{II}$~\cite{Liu2019,Zhang2023} systems, respectively. In particular, the driving scheme of class-$\mathbb{II}$ does not destroy the symmetry of the original static system, hence the Floquet effective Hamiltonian also inherits the same symmetry and can only support a winding number of $\pm 1$~\cite{sky2022}. In contrast, class-$\mathbb{III}$ and class-$\mathbb{IV}$ schemes are composed of static models with different symmetries, and the resulting Floquet Hamiltonian can have different symmetry than the static model. As we will see below, the change of symmetry can lead to a rich phase diagram with exotic topological phases.

\section{Topological phase diagram}
\label{sec:phasediagram}

Topological phase transition in a Floquet system is accompanied by the closing and reopening of quasienergy band gaps. For the periodically driven two-band system considered here, the spectrum is periodic in quasienergy domain and the band touching can take place at quasienergies $\varepsilon_F =0$ and $\pi/T$. Under the two-step quenching protocol, the evolution operator reads $U_T(\bm{q})=e^{-i\bm{h}_2(\bm{q})\cdot \bm{\sigma}T_2}e^{-i\bm{h}_1(\bm{q})\cdot \bm{\sigma}T_1}$ with $T_2=T-T_1$. Using the Euler form of Pauli matrix, $e^{-i\bm{h}\cdot \bm{\sigma}t}=\cos{(|\bm{h}|t)}-i\sin{(|\bm{h}|t)}\bm{h}\cdot \bm{\sigma}/|\bm{h}|$, we can obtain the Floquet effective Hamiltonian $\mathcal{H}_F=\bm{h}_F\cdot \bm{\sigma}$ with $\bm{h}_F=\varepsilon_F \bm{r}/|\bm{r}|$ and
\begin{eqnarray}
		\cos{(\varepsilon_FT)}&=&\cos{|T_1\bm{h}_1|}\cos{|T_2\bm{h}_2|}
		\nonumber \\
		&& -\frac{\bm{h}_1\cdot\bm{h}_2}{|\bm{h}_1|\cdot|\bm{h}_2|}  \sin{|T_1\bm{h}_1|}\sin{T_2\bm{h}_2}, 	\label{eq5} \\
		\bm{r}& =&\frac{\bm{h}_1}{|\bm{h}_1|}\sin{|T_1\bm{h}_1|}\cos{|T_2\bm{h}_2|}
		\nonumber \\
		&& +\frac{\bm{h}_2}{|\bm{h}_2|}\sin{|T_2\bm{h}_2|}\cos{|T_1\bm{h}_1|}
		\nonumber \\
		&&-\frac{\bm{h}_1\times\bm{h}_2}{|\bm{h}_1|\cdot|\bm{h}_2|}\sin{|T_1\bm{h}_1|}\sin{|T_2\bm{h}_2|}.	\label{eq6}
\end{eqnarray}
By substituting the quasienergy $\varepsilon_F =0, \pi/T$ at band touching point into Eq.~(\ref{eq5}), we get the band-touching condition
\begin{eqnarray}
		\frac{\bm{h}_1\cdot\bm{h}_2}{|\bm{h}_1\cdot\bm{h}_2|} = \pm 1, 
		\quad\textrm{and}\quad
		T_1 |\bm{h}_1|\pm T_2|\bm{h}_2| = m\pi
		\label{eq7}
\end{eqnarray}
with $m$ an even (odd) integer for $\varepsilon_F=0$ ($\pi/T$), or
\begin{equation}
	\label{eq8}
	\begin{aligned}
		T_l|\bm{h}_l|=m_l\pi,
	\end{aligned}
\end{equation}
where $m_{l=1,2}$ are integers satisfying $m=m_1+m_2$ being even for $\varepsilon_F=0$ and odd for $\pi/T$.

\begin{figure}[tbp]
	\centering
	\includegraphics[width=1.0\linewidth]{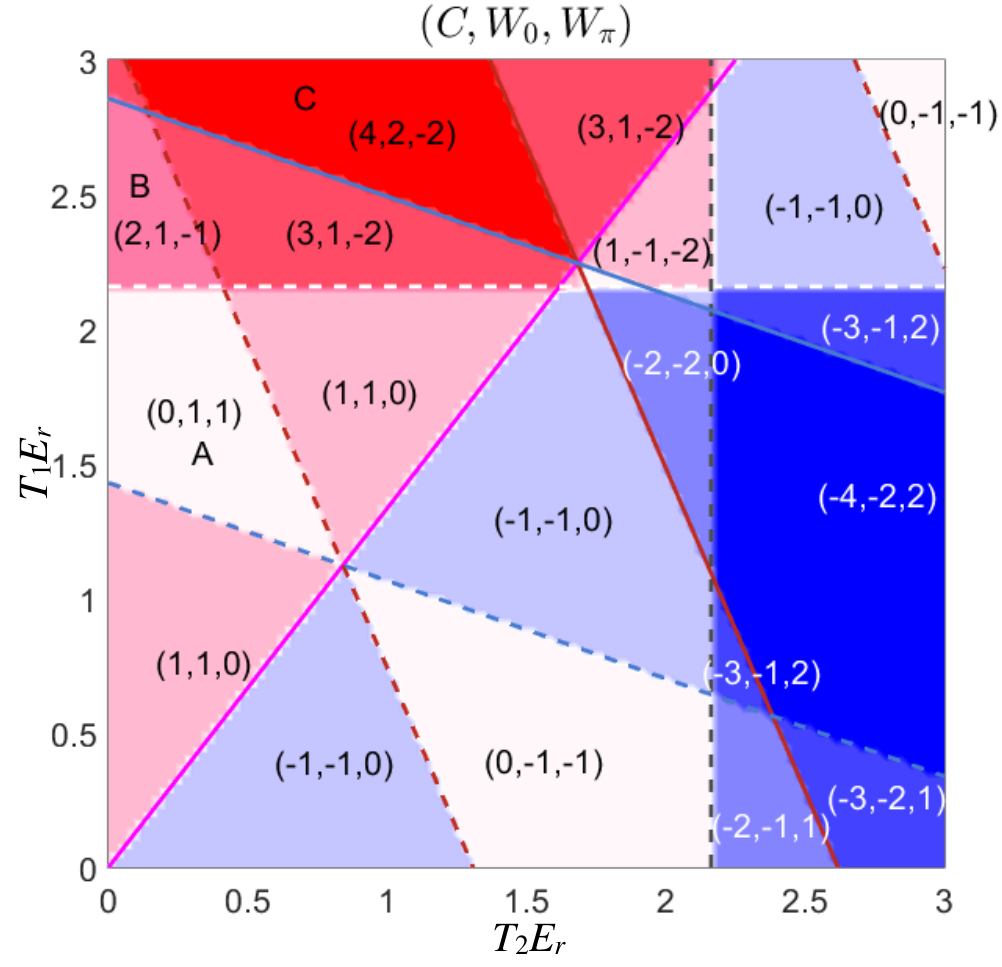}
	\caption{The phase diagram of a Floquet system under class-$\mathbb{III}$ driving scheme. Red, blue and pink lines are the phase boundaries determined by Eq.~(\ref{eq7}). The red lines correspond to the band-touching points at $\Gamma$ point, the blue lines correspond to the $M$ point, and the pink lines to the $X$ point. The white and black lines are the phase boundaries given by Eq.~(\ref{eq8}), and are associated with the band-touching points $\bm{q}^{c2}_{\pm}=(\pm q^{c2}_x,\pi)$ with $q^{c2}_x=\arccos{\frac{m_{z2}+2t^y_{02}}{2t^x_{02}}}$ and $\bm{q}^{c1}_{\pm}=( 0,\pm q^{c1}_y)$ with $q^{c1}_y=\arccos{\frac{m_{z1}-2t^x_{01}}{2t^y_{01}}}$, respectively. The solid lines indicate that the Floquet bands close at $0$ gap ($m$ is even), and the dashed lines indicate that the Floquet bands close at $\pi$ gap ($m$ is odd). The colors used to fill different regions denote different Chern numbers. The parameters of the two instantaneous Hamiltonians considered here are $m_x = m_y=0$, $t^x_{so} = t^y_{so} = 0.2$, $t^x_{01} = t^x_{02} = t^y_{01} = t^y_{02} = 0.4$, $m_{z1} = 0.6$ and $m_{z2} = -0.8$ in unit of $E_r$.}
	\label{fig2}
\end{figure}

The Chern number of the lower quasienergy band can be obtained by $C= W_0-W_{\pi} = \frac{1}{4\pi}\int_{\mathrm{BZ}}d^2\bm{q}\frac{\bm{h}_F}{|\bm{h}_F|^3}\cdot(\partial_{q_x}\bm{h}_F\times \partial_{q_y}\bm{h}_F)$~\cite{Qi2006}, while the Chern number of the upper band satisfies $C^\prime + C = 0$. The Chern number takes an integer value and changes at the phase boundary by 
\begin{equation}
	\label{eq9}
	\Delta C=\sum_{k} \chi(\bm{q}_k),
\end{equation}
where $\chi(\bm{q}_k)=\mathrm{sgn}[(\partial_{q_x}\bm{h}_F\times \partial_{q_y}\bm{h}_F)\cdot \delta \bm{h}_F]|_{\bm{q}=\bm{q}_k}$ and $\bm{q}_k$ is the $k$-th band touching point. By substituting the condition $|r|^2+\cos^2{(\varepsilon_FT)}=1$ guaranteed by the unitarity of $U_T(\bm{q})$, the expression $\bm{h}_F=\varepsilon_F \bm{r}/|\bm{r}|$, and the band touching condition, we get~\cite{Xiong2016}
\begin{equation}
	\label{eq10}
	\chi(\bm{q}_k)=\mathrm{sgn}[(\partial_{q_x}\bm{r}\times \partial_{q_y}\bm{r})\cdot \delta \bm{r}]|_{\bm{q}=\bm{q}_k}.
\end{equation}
In the following, we discuss the phase diagram for class-$\mathbb{III}$ and class-$\mathbb{IV}$ driving schemes with the aid of the aforementioned topological invariants. 

For class-$\mathbb{III}$ driving scheme $\mathcal{H}^x$-$\mathcal{H}^y$, the two instantaneous Hamiltonians can be written without loss of generality as
\begin{eqnarray}
\bm{h}_1 &=& \left( m_x,2t^x_{so}\sin{q_x},m_{z1}-2t^x_{01}\cos{q_x}-2t^y_{01}\cos{q_y} \right),
\nonumber \\
\bm{h}_2 &=& \left( 2t^y_{so}\sin{q_y},m_y,m_{z2}-2t^x_{02}\cos{q_x}-2t^y_{02}\cos{q_y} \right).
\end{eqnarray}
For simplicity, in the following we take $m_x=m_y=0$ and $t^j_{01} =t^j_{02} =$$t_0$. According to Eq.~(\ref{eq7}), we conclude that the band-touching points must locate at the four high symmetry points $\Lambda (\alpha,\beta) =\{\Gamma (0,0), X_1(0,\pi), X_2(\pi,0),M(\pi,\pi) \}$ with $\alpha,\beta=0, \pi$. Using Eq.~(\ref{eq6}), we have $\partial_{q_x}\bm{r}(\alpha,\beta)=2e^{-i\alpha}(-A_x,B_x,0)$, $\partial_{q_y}\bm{r}(\alpha,\beta)=2e^{-i\beta}(B_y,A_y,0)$ and $\bm{r}(\alpha,\beta)=(0,0,r^z)$, where $A_x =-\frac{t^x_{so}h^z_1}{|\bm{h}_1|\cdot|\bm{h}_2|}$ $\sin{|T_1\bm{h}_1|}\sin{|T_2\bm{h}_2|}$, $A_y=\frac{t^y_{so}h^z_2}{|\bm{h}_1|\cdot|\bm{h}_2|}$ $\cdot\sin{|T_1\bm{h}_1|}$$\sin{|T_2\bm{h}_2|}$, $B_x=\frac{t^x_{so}}{|\bm{h}_2|}\sin{|T_2\bm{h}_2|}\cos{|T_1\bm{h}_1|}$, $B_y$ $=\frac{t^y_{so}}{|\bm{h}_1|}\sin{|T_1\bm{h}_1|}\cos{|T_2\bm{h}_2|}$ and $r^z=\sin{(T_1h^z_1+T_2h^z_2)}$. Furthermore, from Eq.~(\ref{eq7}) we know that $|T_1h^z_1+T_2h^z_2|=m\pi$, thus $\delta r^z=(-1)^m\mathrm{sgn}[\delta T_1h^z_1+\delta T_2h^z_2]$. With that, Eq.~(\ref{eq10}) can be reduced to
\begin{equation}
	\label{eq11}
	\chi(\alpha,\beta)_{\mathbb{III}}=(-1)^{m+1}e^{-i(\alpha+\beta)}\mathrm{sgn}[f]\mathrm{sgn}[\delta T_1h^z_1+\delta T_2h^z_2],
\end{equation}
where $f=A_xA_y+B_xB_y$. When $m$ is even or odd, we have $\chi(\alpha,\beta)_{\mathbb{III}}=\Delta W_{0}$ or $-\Delta W_{\pi}$, respectively. Here, $\Delta W_{0}$ and $\Delta W_{\pi}$ are respectively the change of winding numbers at $\varepsilon_{F} = 0$ and $\pi/T$ by crossing the phase transition.

Another kind of band-touching point is determined by Eq.~(\ref{eq8}). From Eqs.~(\ref{eq6}) and (\ref{eq8}), we have
\begin{equation}
	\label{eq12}
	\begin{aligned}
		\partial_{q_j}\bm{r}(\bm{q}^c)=&(-1)^{m_1+m_2} \bigg[T_1\partial_{q_j}|\bm{h}_1(\bm{q}^c)|\frac{\bm{h}_1(\bm{q}^c+\delta q_j)}{|\bm{h}_1(\bm{q}^c+\delta q_j)|}\\
		&+T_2\partial_{q_j}|\bm{h}_2(\bm{q}^c)|\frac{\bm{h}_2(\bm{q}^c+\delta q_j)}{|\bm{h}_2(\bm{q}^c+\delta q_j)|}\bigg],\\
		\delta\bm{r}=&(-1)^{m_1+m_2}\left[\delta T_1\bm{h}_1(\bm{q}^c)+\delta T_2\bm{h}_2(\bm{q}^c)\right],
	\end{aligned}
\end{equation}
where $j=x,y$ and $\bm{q}^c$ denotes the band-touching point satisfying Eq.~(\ref{eq8}). If $m_1\neq 0$ and $m_2\neq 0$, $\bm{h}_l(\bm{q}^c+\delta q_x)=\bm{h}_l(\bm{q}^c+\delta q_y)=\bm{h}_l(\bm{q}^c)$, the vectors $\partial_{q_x}\bm{r}(\bm{q}^c)$, $\partial_{q_y}\bm{r}(\bm{q}^c)$, and $\delta\bm{r}(\bm{q}^c)$ are all in the same plane  spanned by $\bm{h}_1(\bm{q}^c)$ and $\bm{h}_2(\bm{q}^c)$. Thus, one can easily get $\chi(\bm{q}^c)=0$~\cite{Xiong2016}. 
On the other hand, if $m_1=0$, one has $\bm{h}_1(\bm{q}_{\pm}^{c1}+\delta q_x)\neq \bm{h}_1(\bm{q}_{\pm}^{c1}+\delta q_y)$, and the vectors $\partial_{q_x}\bm{r}$, $\partial_{q_y}\bm{r}$ and $\delta\bm{r}$ are not all in the same plane. The band-touching points are then given by $\bm{q}^{c1}_{\pm}=(\alpha,\pm q^{c1}_y)$ with $q^{c1}_y=\arccos{\frac{m_{z1}-2t^x_{01}e^{i\alpha}}{2t^y_{01}}}$. The case of $m_2 = 0$ can be analyzed analogously, leading to the band-touching points $\bm{q}^{c2}_{\pm}=(\pm q^{c2}_x,\alpha)$ with $q^{c2}_x=\arccos{\frac{m_{z2}-2t^y_{02}e^{i\alpha}}{2t^x_{02}}}$.
Combining Eqs.~(\ref{eq10}) and (\ref{eq12}), we obtain 
\begin{eqnarray}
	\label{eq13}
		\chi(\pm q^{c1}_x,\alpha)_{\mathbb{III}} &=& (-1)^{m_2}\mathrm{sgn}[t^x_{so}t^y_{so}t^x_{01}e^{i\alpha}]\delta T_2, \nonumber \\
		\chi(\alpha,\pm q^{c2}_y)_{\mathbb{III}} &=& (-1)^{m_1}\mathrm{sgn}[t^x_{so}t^y_{so}t^y_{02}e^{i\alpha}]\delta T_1.
\end{eqnarray}
When $m_2$ is even (odd), $\chi(\pm q^{c1}_x,\alpha)_{\mathbb{III}}=\Delta W_{0}$ ($-\Delta W_{\pi}$). If $m_1$ is even (odd), we get $\chi(\alpha,\pm q^{c2}_y)_{\mathbb{III}}=\Delta W_{0}$ ($-\Delta W_{\pi}$). To sum up, we can determine the phase diagram. 

\begin{figure}[tbp]
	\centering
	\includegraphics[width=1.0\linewidth]{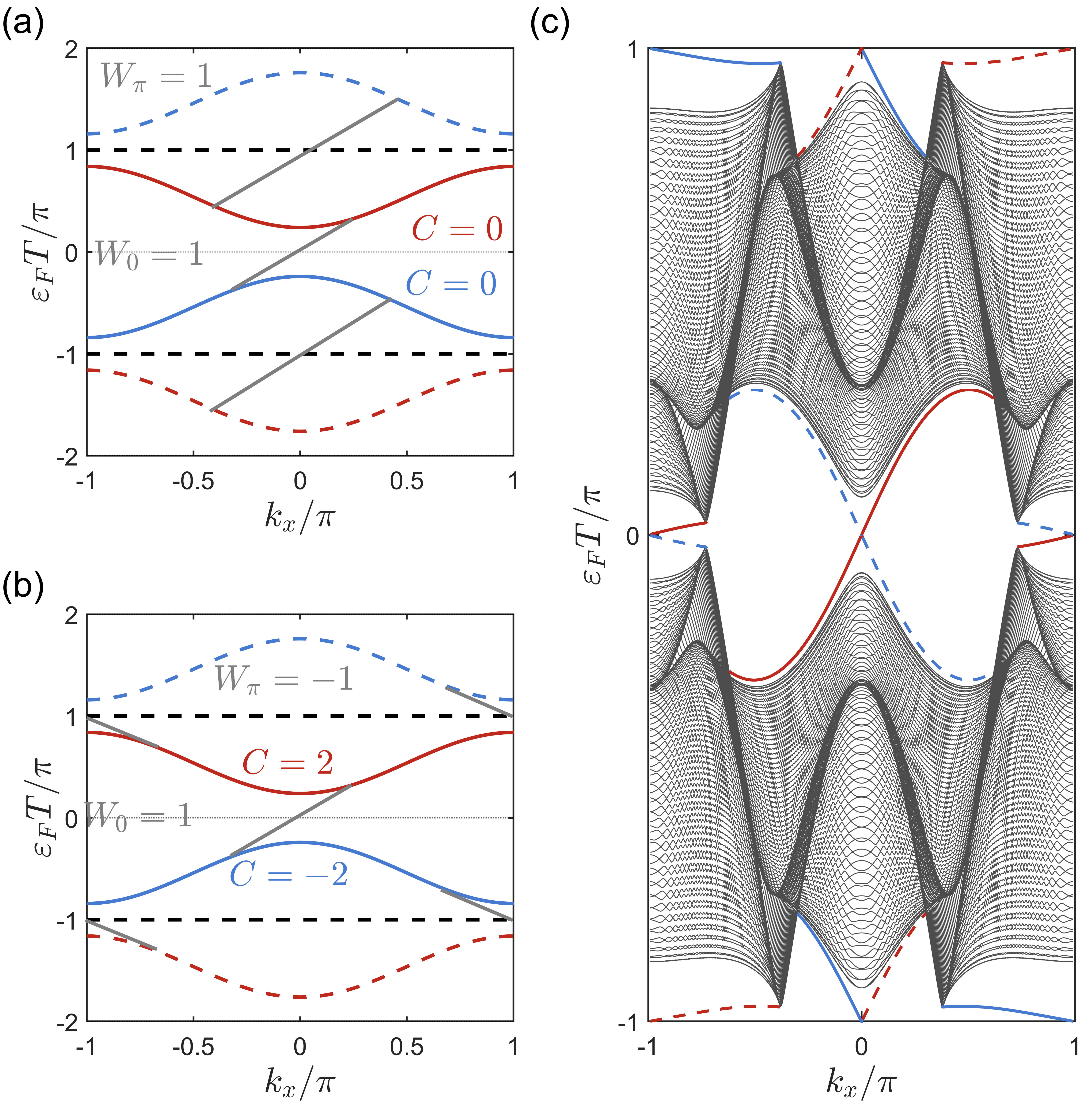}
	\caption{Structure of quasienergy spectrum for (a) AFT phase and (b) the Floquet topological phase with $|C|=2$. (c) Quasienergy spectrum of Floquet topological phase with large winding numbers $|W_{0}|=|W_{\pi}|=2$ under open boundary conditions. The edge modes with positive (negative) chirality are represented in red (blue). Dashed and solid lines represent edge modes located at different boundaries. The durations of two-step driving are $T_1E_r=2.95$ and $T_2E_r=1$, and other parameters are the same as in Fig.~\ref{fig2}.}
	\label{fig3}
\end{figure}

In Fig.~\ref{fig2}, we present a typical example of the topological phase diagram for a class-$\mathbb{III}$ driving system by varying $T_1$ and $T_2$, where the in-plane Zeeman fields $m_x = m_y  = 0$, the spin-orbit coupling intensities $t_{so}^x  = t_{so}^y = 0.2$, the hopping coefficients $t_{01}^x = t_{02}^x = t_{01}^y = t_{02}^y = 0.4$, and the out-of-plane Zeeman fields $m_{z1} = 0.6$ and $m_{z2} = -0.8$. To get connection with cold atom experiment, the recoil energy $E_r=\frac{\hbar^2k^2_0}{2m}$ of the optical lattice is used as the natural energy unit, where $k_0$ is the wave number of lattice beams and $m$ is the atomic mass.
Different topological phases characterized by $(C, W_0, W_\pi)$ are shown. A stark observation is that there exist phases (e.g., the one labeled by `C') with winding numbers as high as $\pm 2$, which thus lead to a Chern number up to $\pm 4$. These exotic phases are induced by the new band-touching points given by the condition Eq.~(\ref{eq8}). In addition, we also find AFT phases (labeled by `A') and large-Chern-number AFT phases (`B').  It is worth noting that here we assume the same hopping rates in the $x$  and $y$ directions, i.e., $t_x^0 = t_y^0$. Under this condition, quasienergy bands touch at the high symmetry points $X_1=(0,\pi)$ and $X_2=(\pi,0)$ simultaneously at the purple line of Fig.~\ref{fig2}, leading to a topological phase transition with winding number change of 2. If the geometric symmetry is broken with $t_x^0 = t_y^0$, this purple topological transition line will split into two lines corresponding respectively to $X_1$ and $X_2$, with a topological number change of only 1.~\cite{Seshadri2019}. On the other hand, a winding number change of 2 can also occur even when the lattice symmetry is broken. For example, at the black or white dashed transition lines of Fig.~\ref{fig2}, Floquet bands touch at $\bm{q}^{c1}_{\pm}$ or $\bm{q}^{c2}_{\pm}$ simultaneously, owing to the symmetry of quasienergy spectra. Thus, the winding number change is always 2 regardless of the geometric symmetry of the underlying lattice.
  
To understand the mechanism of the topological phase with large winding number, it is instructive to compare with class-$\mathbb{II}$ driving scheme under which this exotic phase cannot be found~\cite{sky2022}. For class-$\mathbb{II}$ driving $\mathcal{H}^{\mathrm{AQH}}$-$\mathcal{H}^{\mathrm{AQH}}$, we conclude from Eq.~(\ref{eq7}) that the band-touching points also locate at the high symmetry points $\Lambda(\alpha, \beta)$. Since the instantaneous Hamiltonians both have $C_4$ symmetry with $t^x_{so1}=t^y_{so1}$ and $t^x_{so2}=t^y_{so2}$, we find $A_x=A_y$ and $B_x=B_y$, and $\chi(\alpha,\beta)_{\mathbb{II}}=(-1)^{m+1}e^{-i(\alpha+\beta)} \mathrm{sgn} [\delta T_1h^z_1+\delta T_2h^z_2]$. On the other hand, Eq.~(\ref{eq8}) does not bring other new band-touching points as restricted by $C_4$ symmetry. The symmetry requirement thus limits the choice of topological number, so that the absolute value of the winding number cannot be greater than 1, i.e., $|W_{0/\pi}|\leq 1$. It is also worth noting that class-$\mathbb{II}$ driving also cannot produce the AFT phase with $C=0$ and $W_0=W_{\pi}=1$~\cite{Zhang2023}, as shown in Fig.~\ref{fig3}(a), but can give a phase with large Chern number $C=2$ and $W_0=-W_{\pi}=1$, as shown in Fig.~\ref{fig3}(b). 

\begin{figure}[tbp]
	\centering
	\includegraphics[width=0.9\linewidth]{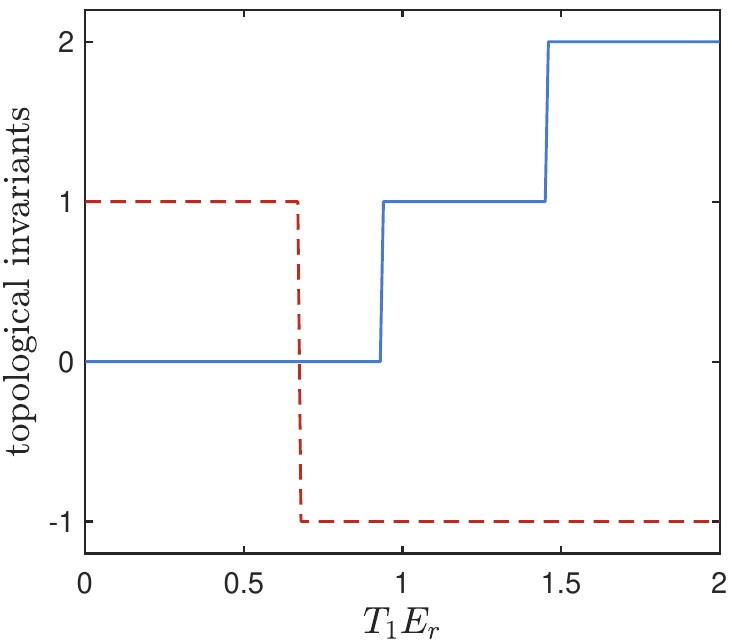}
	\caption{Topological invariants of Floquet topological phases driven by a class-$\mathbb{IV}$ driving scheme ${\cal H}^y$-${\cal H}^{\mathrm{AQH}}$. The red-dashed (blue-solid) line represents the winding number associated with $0$ gap ($\pi$ gap). Here, ${\cal H}^{y}(\bm{q})=2t^y_{so}\sin{q_y}\sigma_x + m_y\sigma_y + (m_{z1}-2t_0^x\cos{q_x} - 2t_0^y\cos{q_y})\sigma_z$ with $(t^y_{so},t_0^x,t_0^y,m_y,m_{z1})=(0.2,0.4,0.4,0,-0.8)E_r$, and ${\cal H}^{\mathrm{AQH}}(\bm{q})=2t_{so}\sin{q_y}\sigma_x + 2t_{so}\sin{q_x}\sigma_y + (m_{z2}-2t_0\cos{q_x} - 2t_0\cos{q_y})\sigma_z$ with $(t_{so},t_0,m_{z2})=(0.2,0.4,0.6)E_r$. The duration of the second phase is fixed at $T_2=0.9/E_r$.}
	\label{fig4}
\end{figure}

For class-$\mathbb{IV}$ driving scheme $\mathcal{H}^y+\mathcal{H}^{\mathrm{AQH}}$ or $\mathcal{H}^x+\mathcal{H}^{\mathrm{AQH}}$, we can employ a similar analysis on band-touching points to map out the phase diagram. As an example, we consider the $\mathcal{H}^y+\mathcal{H}^{\mathrm{AQH}}$ scheme with instantaneous Hamiltonians 
\begin{equation}
	\begin{aligned}
	\bm{h}_1 &= \left( 2t^y_{so}\sin{q_y},m_y,m_{z1}-2t^x_{0}\cos{q_x}-2t^y_{0}\cos{q_y} \right),
     \\
    \bm{h}_2 &= \left(2t_{so}\sin{q_y},2t_{so}\sin{q_x},m_{z2}-2t_{0}\cos{q_x}-2t_{0}\cos{q_y} \right)
	\end{aligned}
\end{equation}
and parameters $m_y=0$, $t_{so}=t^y_{so}$, $t_0^x=t_0^y=t_0$ and $m_{z1}\neq m_{z2}$. The band-touching points given by Eq.~(\ref{eq7}) are still the highly symmetric points $\Lambda (\alpha,\beta)$, and the ones determined by Eq.~(\ref{eq8}) are $\bm{q}^{c}_{\pm}=(\pm q^{c}_x,\alpha)$ with $q^{c}_x=\arccos{\frac{m_{z1}-2t^y_{0}e^{i\alpha}}{2t^x_{0}}}$. The change of topological numbers $\chi(\bm{q}_k)$ caused by the closing and reopening of Floquet bands at band-touching point $\bm{q}_k$ can be obtained from Eq.~(\ref{eq10}). In Fig.~\ref{fig4}, we show the winding numbers of the Floquet topological phases encountered by changing $T_2$ with a fixed $T_1 = 0.9/E_r$. Notice that the winding number of the $\pi$ gap can reach $2$. 

\section{Quasienergy spectrum and topology detection}
\label{sec:detection}

For a static system, a state with large Chern number can be characterized by its novel spectrum of the lowest band, which can host multiple pairs of in-gap edge modes. However, for a Floquet system with infinite numbers of duplicated quasienergy bands, a large Chern number does not guarantee the existence of multiple in-gap modes. Owing to the existence of $\pi$ gap at $\varepsilon_F = \pi/T$ and the identity $C= W_0-W_{\pi}$, it is possible for a phase with Chern number 0 or 2 has a unit winding number and a single pair of in-gap mode for all gaps. In Figs.~\ref{fig3}(a) and \ref{fig3}(b), we show the spectra of an AFT phase (with $C=0$) and a large-Chern-number AFT phase (with $C=2$) under open boundary condition, and find that the two cases both have only one independent edge mode within each gap. As a comparison, the topological phase with large winding number has a very different quasienergy spectrum. In Fig.~\ref{fig3}(c), we present the quasienergy spectrum for a system under class-$\mathbb{III}$ driving, with parameters chosen as in Fig.~\ref{fig2} and durations $T_1E_r=2.95$, $T_2E_r=1$. The choice of parameters is within the red region labeled by `C' with topological numbers $(C, W_0, W_\pi) = (4, 2, -2)$. Obviously there are two pairs of in-gap modes with the same chirality within each gap, in exact correspondence to the value of $W_{0, \pi}$.

\begin{figure}[tbp]
	\centering
	\includegraphics[width=1.0\linewidth]{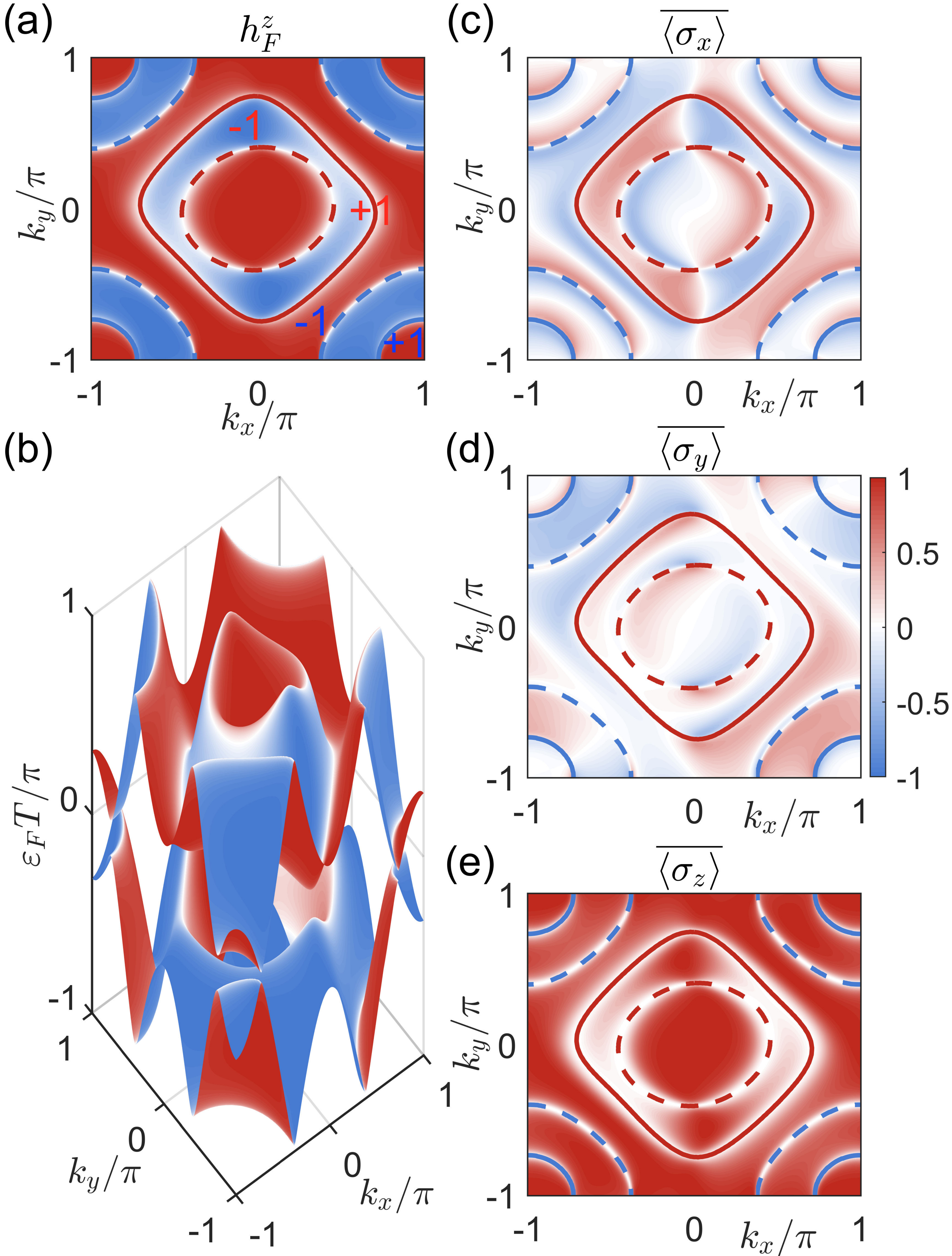}
	\caption{(a) The $z$ component of the equivalent magnetic field and (b) the Floquet bands structure of the `C'-phase. (c-e) The stroboscopic time-averaged spin textures. The solid (dotted) lines represent BISs in the $0$ gap ($\pi$ gap). The parameters used are the same as in Fig.~\ref{fig3}(c).}
	\label{fig5}
\end{figure}

The nontrivial band topology of Floquet topological phase can be characterized by mapping the band inversion surfaces (BISs)~\cite{Zhang2018,Zhang2020}, which are embodied in the dynamical spin pattern of momentum space by quenching from an initially fully-polarized state~\cite{Yi2019,Liang2023}. For the system we are considering, BISs are rings in quasimomentum space satisfying $h^z_F(\bm{q})=0$. In Fig.~\ref{fig5}(a), we plot the $z$ component of the equivalent magnetic field $h_F(\bm{q})$ of the Floquet band for the `C' phase with large winding numbers, where the BISs in $0$ gap ($\pi$ gap) are denoted by solid (dotted) lines. For example, the BIS (dotted line) lying closest to the $\Gamma$ point corresponds to the inner most band touching points located in the $\pi$ gap, while the BIS (solid line) closest to the $M$ point corresponds to the band touching points located in the $0$ gap and at the corners of the Brillouin zone, as can be seen in Fig.~\ref{fig5}(b).

To reveal the BISs in experiments, we propose to start from a fully spin polarized state $\ket{\uparrow}$ and observe the time evolution after quenching to the phase about to be detected~\cite{Zhang2023}. The BISs are formed by the points satisfying \{$\bm{q}|\overline{\langle\sigma_i\rangle}=0$, $\forall i$\}, where $\overline{\langle\sigma_i\rangle}$ is the stroboscopic time-averaged spin texture with
\begin{equation}
	\label{eq14}
	\overline{\langle\sigma_i\rangle}=\lim_{N\rightarrow\infty}\frac{1}{N}\sum^N_{n=0}\langle\sigma_i(\bm{q},t=nT)\rangle.
\end{equation}
For the `C' phase of most interest, we show the stroboscopic time-averaged spin textures in Figs.~\ref{fig5}(c)-\ref{fig5}(e), from which four BISs, two $0$ BISs and two $\pi$ BISs, can be clearly observed. The topological number $\nu$ of each BIS is determined by the winding of the dynamic field $\bm{g}(\bm{q})=(g_y,g_x)$ along the BIS~\cite{Zhang2020}, where $g_i=-\partial_{q_{\bot}}\overline{\langle\sigma_i\rangle}/N_q$ with $q_{\bot}$ being the momentum perpendicular to the BIS and pointing from $h_F^z<0$ to $h_F^z>0$, and $N_q$ denotes the normalization factor. The winding numbers $W_{0,\pi}$ associated with different gaps can be calculated by summing over $\nu$'s of the BISs located therein. Therefore, through the stroboscopic time-averaged spin textures, we can conclude that the topological numbers of both $0$ BISs are $+1 $, and the topological numbers of both $\pi$ BISs are $-1$, thus leading to the topological numbers of the `C' phase as $(C,W_0,W_{\pi}) = (4,2,-2) $.

\section{Conclusion}
\label{sec:summary}

We propose to realize novel Floquet topological phases with large winding number (higher than 1) in periodically driven systems. By considering a two-step driving scheme in a two-dimensional model with spin-orbit coupling and Zeeman field, we calculate the quasienergy spectrum and analytically obtain the locations of band-touching points for different choices of instantaneous Hamiltonian parameters. We find that if the driving scheme is composed of instantaneous Hamiltonians with different spatial symmetries, new Floquet topological phases with winding number as large as $\pm 2$ can be observed. The presence of such exotic phases are rooted from the different symmetry groups of the effective Floquet Hamiltonian, and is unique for periodically driven systems without any static counterpart. Further, we show that the topology of such Floquet topological phases can be characterized experimentally by detecting the stroboscopic time-averaged spin textures in quench dynamics. We stress that the model considered can be readily realized in cold atoms trapped in optical Raman lattice~\cite{Wu2016,Yi2019,Liang2023,Sun2018,S-Y2018}.

\section{Acknowledgments}
This work is supported by the National Natural Science Foundation of China (Grants No.~12074427, No.~12074428 and No.~92265208), and the National Key R\&D Program of China (Grants No.~2018YFA0306501 and No.~2022YFA1405301).


\end{document}